\documentclass[aps,twocolumn,longbibliography] {revtex4-1}
\usepackage{amsmath,graphicx,amsfonts}

\usepackage{dcolumn}
\newcolumntype{L}{>{$}l<{$}}
\newcolumntype{Q}{>{$}c<{$}}

\usepackage{afterpage}
\usepackage{makecell}

\usepackage{times}
\usepackage[colorlinks=true,linkcolor=blue,urlcolor=blue,citecolor=blue, breaklinks=true,hypertexnames=false]{hyperref}

\usepackage{amssymb}
\usepackage{bbm}

\usepackage[utf8]{inputenc}
\DeclareUnicodeCharacter{03BC}{\mbox{$\mu$}}

\DeclareUnicodeCharacter{03B4}{$\delta$}
\DeclareUnicodeCharacter{2009}{-}

\usepackage[english]{babel}
\addto\captionsenglish{}
\addto\captionsenglish{}

\makeatletter
\def\bbl@set@language#1{%
	\edef\languagename{%
		\ifnum\escapechar=\expandafter`\string#1\@empty
		\else\string#1\@empty\fi}%
	\@ifundefined{babel@language@alias@\languagename}{}{%
		\edef\languagename{\@nameuse{babel@language@alias@\languagename}}%
	}%
	\select@language{\languagename}%
	\expandafter\ifx\csname date\languagename\endcsname\relax\else
	\if@filesw
	\protected@write\@auxout{}{\string\select@language{\languagename}}%
	\bbl@for\bbl@tempa\BabelContentsFiles{%
		\addtocontents{\bbl@tempa}{\xstring\select@language{\languagename}}}%
	\bbl@usehooks{write}{}%
	\fi
	\fi}
\newcommand{\DeclareLanguageAlias}[2]{%
	\global\@namedef{babel@language@alias@#1}{#2}%
}
\makeatother

\DeclareLanguageAlias{en}{english}

\makeatletter
\def\@bibdataout@aps{%
	\immediate\write\@bibdataout{%
		@CONTROL{%
			apsrev41Control%
			\longbibliography@sw{%
				,author="08",editor="1",pages="1",title="0",year="1"%
			}{%
				,author="08",editor="1",pages="1",title="",year="1"%
			}%
		}%
	}%
	\if@filesw \immediate \write \@auxout {\string \citation {apsrev41Control}}\fi 
}
\makeatother

\date\today

\newcommand\trick[1]{}

\def\TITLE{Local Properties of the Rapidity Distribution in the Lieb--Liniger Model}

\begin{document}
\title{\TITLE}
	\author{Mi\l{}osz Panfil$^1$ and Zoran Ristivojevic$^2$}
	\affiliation{$^1$Faculty of Physics, University of Warsaw, Pasteura 5, 02-093 Warsaw, Poland}
	\affiliation{$^2$Laboratoire de Physique Th\'{e}orique, Universit\'{e} de Toulouse, CNRS, UPS, 31062 Toulouse, France}
	
\begin{abstract}
We study the rapidity distribution in the Lieb--Liniger model and derive exact relations for its derivatives at the Fermi level. The latter enables us to treat analytically the free energy of the system at low temperatures and arbitrary interactions. We calculated the leading-order correction to the well-known result obtained using conformal field theory. In contrast to the leading-order term controlled by the sound velocity or the Luttinger liquid parameter, the new term is controlled by an additional dimensionless parameter. We calculated its series expansions in the limiting cases of weak and strong interactions. Our results are generalized to other Galilean-invariant integrable systems. 
\end{abstract}
\maketitle
	
\textit{Introduction.---}The system of one-dimensional bosons with short-range interactions is described by the Lieb--Liniger model. It admits an exact solution in terms of the Bethe ansatz \cite{lieb_exact_1963}, enabling studies of the interaction effects in an exact way \cite{korepin,gaudin}. Experimental realizations \cite{paredes_tonksgirardeau_2004,kinoshita_observation_2004} further strengthen physical importance of the model, which has become a cornerstone of our understanding of correlation effects and many-body dynamics in one-dimensional quantum physics.

The Bethe-ansatz wavefunction is parametrized by the set of rapidities. In the thermodynamic limit, the rapidity distribution is the central object that controls many quantities in the Lieb--Liniger model. For example,  its zeroth and second moments lead to the particle density and the energy, respectively, while higher moments determine expectation values of conserved charges and various local correlation functions \cite{cheianov_exact_2006,pozsgay_local_2011,bastianello_exact_2018,ristivojevic_method_2022}. Similarly, the rapidity distribution controls the large-distance and long-time asymptotes of the correlation functions \cite{shashi_exact_2012,kitanine_form_2012,de_nardis_exact_2016}. It is also of prime importance to non-equilibrium dynamics \cite{caux_time_2013,caux_quench_2016,castro-alvaredo_emergent_2016,bertini_transport_2016}. Furthermore, the ground-state value of the rapidity distribution at the edge determines the Luttinger liquid parameter, which automatically gives the sound velocity as a consequence of Galilean invariance \cite{haldane_effective_1981}. Interestingly, the rapidity distribution has a direct relation with seemingly unrelated capacitance of a circular capacitor \cite{gaudin}. Finally, the rapidity distribution can be understood as the asymptotic momentum distribution after the free expansion of the Bose gas {in one dimension}, which was used in the recent experiments \cite{wilson_observation_2020,dubois_probing_2024} to directly measure it.

The thermodynamics of the model can also be calculated formally exactly using the Yang--Yang formalism \cite{yang_thermodynamics_1969}. Moreover the thermodynamics was directly probed in experiments \cite{van_amerongen_yang-yang_2008,jacqmin_sub-poissonian_2011,vogler_thermodynamics_2013}. Despite the exact solution, extracting simple analytical forms of relevant quantities valid without limitation on the interaction strength is a formidable task. A notable exception is the leading low-temperature result for the free energy, which was obtained in the framework of conformal field theory \cite{affleck_universal_1986,blote_conformal_1986}. It relies on the linear spectrum of low-energy excitations. This approach should be contrasted to the others, based on the Bethe ansatz \cite{guan_polylogs_2011} or the effective quasiparticle picture \cite{kerr_analytic_2024,de_rosi_beyond-luttinger-liquid_2019}, that treat the nonlinear spectrum but are limited to weak or strong interactions. 
	
In this paper we overcome these difficulties by developing the exact description of local properties of the rapidity distribution of the system. Combined with the Yang--Yang theory, the latter enables us to find the exact result for the low-temperature free energy. It is valid at any interaction and can be understood as the leading correction to the conformal field theory result.

\textit{Partial differential equation for the rapidity distribution.---}Consider bosons of the mass $m$ in one dimension with repulsive interaction potential $V(x)=(\hbar^2 c/m)\delta(x)$. Their ground-state rapidity distribution $\rho(k,Q)$ satisfies the Lieb equation \cite{lieb_exact_1963}
\begin{align}\label{eq:LIE}
\rho(k,Q)+\frac{1}{2\pi}\int_{-Q}^{Q}dq\;\! \theta'(k-q)\rho(q,Q)=\frac{1}{2\pi}.
\end{align}
Here $\theta(k)=-2\arctan(k/c)$ is the two-particle scattering phase shift and $Q$ is the Fermi rapidity, which denotes the highest occupied rapidity in the ground state. Instead of dealing with the integral equation (\ref{eq:LIE}), it can be differentiated leading to \cite{petkovic_spectrum_2018,ristivojevic_exact_2023}
\begin{align}\label{eq:PDE}
\left(\frac{\partial^2}{\partial Q^2}-\frac{2}{\rho(Q,Q)}\frac{d\rho(Q,Q)}{dQ}\frac{\partial}{\partial Q}-\frac{\partial^2}{\partial k^2}\right)\rho(k,Q)=0.
\end{align}
The partial differential equation (\ref{eq:PDE}) is very convenient to study the local properties of $\rho(k,Q)$ for $k$ in the vicinity of $Q$.
	
Partial derivatives of the rapidity distribution {$\rho^{(j,l)}(Q,Q)$, where
$\rho^{(j,l)}(k,Q)={\partial^{j+l} \rho(k,Q)}/{\partial k^j \partial Q^l}$},
are generally unknown. Here it will be first shown that they satisfy certain linear differential equations and at a second step the latter equations will be solved. As a starting point we use Eq.~(\ref{eq:PDE}) and the expressions for the total derivative of $\rho(Q,Q)$,
\begin{align}\label{eq:totalderivative}
\frac{d^j}{dQ^j}\rho(Q,Q) -\sum_{l=0}^{j}\binom{j}{l}\rho^{(l,j-l)}(Q,Q)=0.
\end{align}
In the following we assume that $\rho(Q,Q)$ is a known function. As a consequence, the first term in the left-hand side of Eq.~(\ref{eq:totalderivative}) is also known. Straightforward calculation shows that $\rho^{(1,0)}(Q,Q)$ satisfies the equation  
\begin{align}\label{eq:Fho10}
\frac{d}{dQ}\left(\frac{\rho^{(1,0)}(Q,Q)}{\rho(Q,Q)}\right)=R(Q),
\end{align}
where we have introduced
\begin{align}
R(Q)=\frac{\ddot{\rho}(Q,Q)}{2\rho(Q,Q)}-\left(\frac{\dot{\rho}(Q,Q)} {\rho(Q,Q)}\right)^2.
\end{align}
Here and in the following the dots over $\rho(Q,Q)$ denote the total derivatives, i.e., $\dot\rho(Q,Q)=d\rho(Q,Q)/dQ$, $\ddot\rho(Q,Q)=d^2\rho(Q,Q)/dQ^2$, etc. Equation (\ref{eq:Fho10}) has the form of a first-order linear differential equation for the function $\rho^{(1,0)}(Q,Q)/\rho(Q,Q)$, since the right-hand side, $R(Q)$, is by assumption a known function. We note that once the partial derivative $\rho^{(1,0)}(Q,Q)$ is evaluated,  $\rho^{(0,1)}(Q,Q)$ follows directly from Eq.~(\ref{eq:totalderivative}).
	
The second partial derivative obeys 
\begin{align}\label{eq:Fho20}
\frac{d}{dQ}\left(\frac{\rho^{(2,0)}(Q,Q)}{\rho(Q,Q)}\right)=\frac{1}{2}\frac{dR(Q)}{dQ}+R(Q)\frac{\rho^{(1,0)}(Q,Q)}{\rho(Q,Q)}.
\end{align}
Using the relation (\ref{eq:Fho10}), Eq.~(\ref{eq:Fho20}) can be integrated, yielding
\begin{align}\label{eq:Fho20solution}
\frac{\rho^{(2,0)}(Q,Q)}{\rho(Q,Q)}	={}&\frac{1}{2} \frac{d}{dQ}\left(\frac{\rho^{(1,0)}(Q,Q)}{\rho(Q,Q)}\right)\notag\\
		&+\frac{1}{2}\left(\frac{\rho^{(1,0)}(Q,Q)}{\rho(Q,Q)}\right)^2.
\end{align}
Here the integration constant is set to zero accounting for the limiting case of strong interactions where Eq.~(\ref{eq:Fho20}) can be explicitly solved. The expression (\ref{eq:Fho20solution}) is a remarkable result as it shows that the second partial derivative $\rho^{(2,0)}(Q,Q)$ is not independent, but can be expressed in terms of $\rho^{(1,0)}(Q,Q)$. The remaining second derivatives can be now straightforwardly obtained. The result for $\rho^{(0,2)}(Q,Q)$ follows from Eq.~(\ref{eq:PDE}), which can then be used in Eq.~(\ref{eq:totalderivative}) to obtain $\rho^{(1,1)}(Q,Q)$.

Equation (\ref{eq:Fho20}) can be derived as follows. We form a linear combination that consists of $d\left({\rho^{(2,0)}(Q,Q)}/{\rho(Q,Q)}\right)/dQ$ with the coefficient $1$, the left-hand sides of Eq.~(\ref{eq:PDE}) and its $k$- and $Q$-derivatives, all three taken at $k=Q$, as well as the left-hand side of Eq.~(\ref{eq:totalderivative}) taken at $j=1,2,3$. The linear combination involves six parameters that can take arbitrary values as they multiply the expressions that are formally equal to zero. We then impose zero coefficients in front of the terms involving all the partial derivatives $\rho^{(j,l)}(Q,Q)$ but $\rho^{(1,0)}(Q,Q)$. As it turns out, the solution can be found, which once returned to the linear combination yields the right-hand side of Eq.~(\ref{eq:Fho20}). The latter procedure can be straightforwardly extended to find the differential equation for the derivative of $\rho^{(j,0)}(Q,Q)/\rho(Q,Q)$ with $j>2$, see Appendix.
	
\textit{New dimensionless parameters.---}Instead of dealing with the partial derivatives of the rapidity distribution it is more convenient to define the corresponding family of dimensionless parameters. Introducing
\begin{align}
	&K_0(q)=4\pi^2\rho(q,Q)^2,\\
	&K_j(q)=-\pi n\frac{d K_{j-1}(q)}{dq},\quad j=1,2,\cdots,
\end{align}
we define a family of parameters by
\begin{align}\label{eq:Kll}
	&K=K_0(Q),\\
\label{eq:Kj}
	&K_j=K_j(Q),\quad j=1,2,\cdots.	
\end{align}	
Here $n$ denotes the particle density and $K$ is the Luttinger liquid parameter. It can be obtained from the value of the rapidity distribution at the Fermi rapidity \cite{korepin}. The following two parameters of the family are
\begin{align}\label{eq:K1}
	&K_1=-2\pi n K \frac{\rho^{(1,0)}(Q,Q)}{\rho(Q,Q)},\\
\label{eq:K2}
&K_2=\frac{K_1^2}{2K}+2\pi^2 n^2 K \frac{\rho^{(2,0)}(Q,Q)}{\rho(Q,Q)}.
\end{align}
Due to the connection (\ref{eq:Fho20solution}), $K_2$ can be explicitly expressed through $K$ and $K_1$. The derivatives with respect to $Q$ can be transformed to be with respect to $\gamma=c/n$ using the rule
\begin{align}\label{eq:ruleQgamma}
\frac{\partial}{\partial Q}=-\frac{K\gamma^2}{\pi c}\frac{\partial}{\partial\gamma},
\end{align}
where $c$ assumes a constant value. This yields
\begin{align}
\label{eq:K2K1}
K_2=\frac{3K_1^2}{4K}+\frac{K K_1}{2}+\frac{\gamma}{2}(K_1' K-K_1 K').
\end{align}
Similarly as $K$, the parameters $K_1$ and $K_2$ are dimensionless and only depend on $\gamma$. We use the notation $K'={dK}/{d\gamma}$ and similarly $K_1'={dK_1}/{d\gamma}$.

\textit{Results in terms of power series.---}The Luttinger liquid parameter $K(\gamma)$ has a power-series expansion in the regimes of weak, $\gamma\ll 1$, and strong interactions, $\gamma\gg 1$. From the relation (\ref{eq:Kll}) it thus follows that the right-hand sides of Eqs.~(\ref{eq:Fho10}) and (\ref{eq:Fho20}) also have the power-series form. Since the latter differential equations are linear, they can be solved provided there is information about one integration constant. The integration constants can be found from the behavior of $\rho(Q+q,Q)$ at $0<q\ll Q$. This can be obtained by solving Eq.~(\ref{eq:LIE}) in the vicinity of the Fermi rapidity. In this case Eq.~(\ref{eq:LIE}) reduces to an equation of  Wiener--Hopf type which was solved in Refs.~\cite{popov_theory_1977,pustilnik_low-energy_2014,pustilnik_fate_2015}. At the leading order in $\gamma\ll 1$, the final result is given by
\begin{align}
\rho(Q+q,Q)=\frac{1}{2\sqrt\pi \gamma^{1/4}} f\left(\frac{2\pi}{c} q\right),
\end{align}
where
\begin{align}\label{eq:fLLWH}
f(y)=\frac{1}{\sqrt{2}\pi^{3/2}}\int_0^\infty dz \frac{\sin(2\pi z)\Gamma(z)e^{-z(\ln z-1+y)}}{\sqrt{z}}.
\end{align}
The expression for the derivatives of the rapidity distribution at the Fermi rapidity are then
\begin{subequations}\label{eq:FholeadingorderWH}
\begin{align}\label{eq:Fholeadingorder}
\rho^{(j,0)}(Q,Q)=\frac{1}{2\sqrt{\pi}\gamma^{1/4}}\left(\frac{2\pi}{c}\right)^j f^{(j)}(0).
\end{align}
Equation (\ref{eq:Fholeadingorder}) is the leading-order result for all the derivatives at $\gamma\ll 1$. The values of the function (\ref{eq:fLLWH}) and of its first two derivatives at zero are given by \cite{Note1} \footnotetext{Interestingly, the numbers in Eq.~(\ref{eq:fderivatives}) are identical to the coefficients appearing in Stirling's series for the Gamma function, $\Gamma(z)$, written in the form
\begin{align*}
\frac{\Gamma(-z)\sqrt{-z}e^{z(\ln(-z)-1)}}{\sqrt{2\pi}}={}&1-\frac{1}{12z}+\frac{1}{288z^2}+\mathcal{O}((-z)^{-3}).
\end{align*}\protect\trick.
} 
\begin{gather}\label{eq:fderivatives}
f(0)=1,\quad f'(0)=-\frac{1}{12},\quad	f''(0)=\frac{1}{288}.
\end{gather}	
\end{subequations}
Together with our differential equations, the information contained in Eq.~(\ref{eq:FholeadingorderWH}) suffices to obtain the whole power-series form for all the derivatives $\rho^{(j,0)}(Q,Q)$ in the regime $\gamma\ll 1$. In the opposite regime of strong interactions, from a simple analysis of Eq.~(\ref{eq:LIE}) it follows that $\rho^{(j,0)}(Q,Q)=0$ for $j>0$ at $\gamma\to\infty$, which can be used to fix the integration constants. We finally notice that the derivatives with respect to $Q$ that enter Eqs.~(\ref{eq:Fho10}) and (\ref{eq:Fho20}) should be expressed in terms of $\gamma$ according to the rule (\ref{eq:ruleQgamma}).

The above procedure complemented with the known result for $K(\gamma)$ \cite{ristivojevic_conjectures_2019,ristivojevic_excitation_2014} leads to the power-series results for the derivatives $\rho^{(j,0)}(Q,Q)$. They can be then used to find the parameters $K_j$. At weak interactions, $\gamma\ll 1$, we obtain
{\begin{align}
{}&K_1=\frac{\pi^3}{3\gamma^{3/2}}-\frac{7\pi^2}{24\gamma} -\frac{3\pi}{64\sqrt\gamma}  +\frac{1+2\zeta(3)}{256} \notag\\
&+\frac{89+153\zeta(3)}{12288\pi}\sqrt\gamma +\frac{45\zeta(5)+213\zeta(3)-9}{16384\pi^2}\gamma+\mathcal{O}(\gamma^{3/2}).
\end{align}}
The parameter $K_2$ can then be easily found from Eq.~(\ref{eq:K2}). We note that $K$, $K_1$, and $K_2$ are on the same order in the regime of weak interactions. At strong interactions, $\gamma\gg 1$, we find
\begin{gather}
K_1=\frac{8\pi^2}{\gamma^3}-\frac{32\pi^4}{\gamma^5} +\mathcal{O}(\gamma^{-6}),
\end{gather}
while $K_2\sim 1/\gamma^2$. Therefore, $K_1$ and $K_2$ tend to zero as the interaction strength is increased, which should be contrasted to $K$ that reaches one. 
	
\textit{The free energy.---}Previous considerations enable us to study the thermodynamics. The free energy per particle of the system in the canonical ensemble in units of $\epsilon=\hbar^2 n^2/2m$ has the universal low-temperature form 
\begin{align}\label{eq:f}
f(\gamma,\tau)=e(\gamma)-\frac{K}{12}\tau^2+C_4\tau^4+\mathcal{O}(\tau^6).
\end{align}
Here $\tau=T/\epsilon$, where $T$ is the temperature. At $T=0$, the free energy reduces to the ground-state energy of the Lieb--Liniger model, $E_0=N\epsilon\;\! e(\gamma)$, where $N$ is the number of particles. The leading correction is quadratic in $\tau$ and has the well-known form initially obtained from the conformal field theory arguments \cite{blote_conformal_1986,affleck_universal_1986}. It originates from the linear spectrum of elementary excitations at low momenta characterized by the sound velocity $v$. In Eq.~(\ref{eq:f}), the leading correction is expressed in terms of $K$ rather than $v$ using the connection $mvK=\pi\hbar n$, valid in Galilean-invariant systems \cite{haldane_effective_1981}. The leading nontrivial term in Eq.~(\ref{eq:f}) is the one proportional to $\tau^4$. It arises from the spectrum curvature and was previously unknown. Here we have found that it is given by
\begin{align}\label{eq:C4}
C_4={}&\frac{29K^2K_1}{5760\pi^2}+\frac{7\gamma KK'K_1}{5760\pi^2}+\frac{\gamma K^2K_1'}{1152\pi^2}-\frac{K^4}{180\pi^2}\notag\\
&-\frac{49\gamma K^3 K'}{5760\pi^2} -\frac{\gamma^2K^2 K'^2}{1536\pi^2} -\frac{17 \gamma^2K^3K''}{11520\pi^2}.
\end{align}
Equations (\ref{eq:f}) and (\ref{eq:C4}) comprise the main result of this paper. The coefficient (\ref{eq:C4}) is exact. It is expressed only in terms of the dimensionless parameters $K$ and $K_1$. This results shows that the local properties of the rapidity distribution at the Fermi rapidity, i.e., its value $\rho(Q,Q)$ and the first partial derivative $\rho^{(1,0)}(Q,Q)$, fully determine the quartic term in the free energy.  
	
Let us discuss values of $C_4$ in the limiting cases. At the leading order in $\gamma\ll 1$, we have $K={\pi}/{\sqrt\gamma}$. Therefore, only the terms that involve $K_1$ in Eq.~(\ref{eq:C4}) give rise to the leading term given by $C_4={\pi^3}/{960\gamma^{5/2}}$. Accounting for more terms in the expansion we find
{\begin{align}\label{eq:C4small}
C_4={}&\frac{\pi^3}{960\gamma^{5/2}}-\frac{7\pi^2}{2304\gamma^2} -\frac{137\pi}{36864\gamma^{3/2}} -\frac{69-2\zeta(3)}{20480\gamma}\notag\\
&- \frac{4573+957\zeta(3)}{2359296\pi\sqrt\gamma}- \frac{5575+6493\zeta(3)-315\zeta(5)}{7864320\pi^2}\notag\\
&+\mathcal{O}(\sqrt\gamma).
\end{align}}
In sharp contrast to the case of weak interactions, at strong ones, $\gamma\gg 1$, the summands of Eq.~(\ref{eq:C4}) that involve $K_1$ enter starting from the third subleading term. In practice they can thus be often neglected. The leading order term $C_4=-{1}/{180\pi^2}$ is obvious from the structure of Eq.~(\ref{eq:C4}), while further terms in the expansion are given by
{\begin{align}\label{eq:C4large}
{}&C_4=-\frac{1}{180\pi^2}-\frac{1}{15\pi^2\gamma}-\frac{1}{3\pi^2\gamma^2} +\frac{13\pi^2-120}{135\pi^2\gamma^3}\notag\\
&+\frac{26\pi^2-60}{45\pi^2\gamma^4} -\frac{74\pi^4-260\pi^2+240}{225\pi^2\gamma^5} +\mathcal{O}(\gamma^{-6}).
\end{align}}
Since $C_4$ is positive at weak interactions and negative at strong ones, the sign of $C_4$ changes as the interaction strength $\gamma$ is increased. We have found that $C_4$ nullifies at $\gamma\approx 0.606$. The plot of $C_4$ is shown in Fig.~\ref{fig:C4}. {In Appendix we give additional details about the numerical results and its comparison to Eqs.~(\ref{eq:C4small}) and (\ref{eq:C4large}).}

\begin{figure}
\includegraphics[width=\columnwidth]{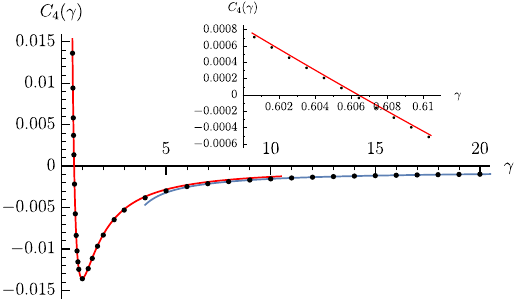}
\caption{Plot of the coefficient $C_4$. The dots represent numerically exact values and the curves are obtained from the analytical results (\ref{eq:C4small}) and (\ref{eq:C4large}). {In the crossover region corresponding to $\gamma$  around $5$, the relative difference between the exact values and the values predicted by Eqs.~(\ref{eq:C4small}) and (\ref{eq:C4large}) reaches its maximum that is around  $0.06$ in absolute value. The inset shows the behavior of $C_4$ in the region where it nullifies.}}\label{fig:C4}
\end{figure}

\begin{figure}
\includegraphics[width=\columnwidth]{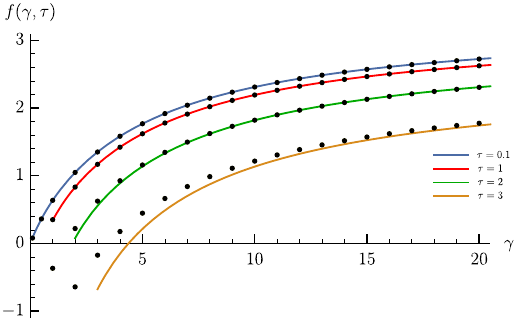}
\caption{The dimensionless free energy per particle $f(\gamma,\tau)$ obtained from the analytical expression (\ref{eq:f}) is represented by the curves and the one obtained by numerically solving the Yang--Yang equation is shown by the dots. Four distinct values of $\tau$ are considered: $0.1$, $1$, $2$, and $3$. For a fixed value of $\gamma$, the function $f(\gamma,\tau)$ decreases when $\tau$ increases. This is expected as the entropy of the system increases with the temperature. The agreement between Eq.~(\ref{eq:f}) and numerically exact values is perfect for $\tau$ smaller than 1 and $\gamma$ larger than $\tau$. On the resolution of the plot, the curve for $\tau=0$ would be indistinguishable from the plotted one for $\tau=0.1$.}\label{fig:freeenergy}
\end{figure}

The limiting cases of $C_4$ can be understood in terms of simple physics. At weak interactions, the free energy of the system can be calculated by studying the statistical mechanics of bosonic quasiparticles with Bogoliubov spectrum. It yields the result (\ref{eq:C4small}) taken at the leading order \cite{de_rosi_beyond-luttinger-liquid_2019}. Interestingly, Bogoliubov spectrum is not the correct form of the quasiparticle spectrum at smallest momenta, as it is replaced by the fermionic one \cite{pustilnik_low-energy_2014,petkovic_spectrum_2018}. However, such picture with bosonic quasiparticles is sufficient in order to reproduce the leading-order coefficient in front of $\tau^4$ power in the free energy \footnote{Recall that the ground-state energy of the Lieb--Liniger model at the leading and the subleading order is correctly reproduced by the same approach with Bogoliubov quasiparticles \cite{lieb_exact_1963}.}. In the limiting case of strong interactions, by calculating the free energy of the gas of noninteracting fermions we recovered the leading order term in Eq.~(\ref{eq:C4large}). Nevertheless, the expression (\ref{eq:C4}) applies at any interaction. In Fig.~\ref{fig:freeenergy} we show the dimensionless free energy per particle for different values of the temperature.
	
\textit{The derivation of Eq.~(\ref{eq:f}).---}The thermodynamic properties of the system can be calculated exactly. They follow from the Yang--Yang equation \cite{yang_thermodynamics_1969}
\begin{align}\label{eq:YY}
\bar{\mathcal{E}}(k)=-\bar\mu+\frac{\hbar^2 k^2}{2m}+ \frac{1}{2\pi}\int_{-\infty}^{+\infty} dq\;\! \frac{\theta(q-k)\bar{\mathcal{E}}'(q)}{1+e^{\bar{\mathcal{E}}(q)/T}}.
\end{align}
The solution $\bar{\mathcal{E}}(k)$ of Eq.~(\ref{eq:YY}) depends on the chemical potential $\bar\mu$ and the temperature $T$. It enables one to calculate the pressure of the system in the grand-canonical ensemble,
\begin{align}\label{eq:pressureYY}
\bar P= \frac{1}{2\pi}\int_{-\infty}^{+\infty} dq\;\! \frac{q\;\! \bar{\mathcal{E}}'(q)}{1+e^{\bar{\mathcal{E}}(q)/T}}.
\end{align}
The free energy of the system in the canonical ensemble is then given by $F=-\bar PL+\bar\mu N$, where $L$ is the system size. 
	
From a basic analysis it follows that $\bar{\mathcal{E}}(k)$ is an even real function, monotonically increasing for $k>0$. It thus nullifies at some $k=\bar Q$,
\begin{align}\label{eq:Energy-Q-condition}
\bar{\mathcal{E}}(\bar Q)=0.
\end{align}
In the zero-temperature limit, the denominator in the integral of Eq.~(\ref{eq:YY}) makes the boundary of integration finite, $Q$, corresponding the the Fermi rapidity. At low temperatures, Eq.~(\ref{eq:YY}) admits a series expansion in even powers of temperature, $\bar{\mathcal{E}}(k)=\mathcal{E}(k)+T^2 \mathcal{E}_2(k)+T^4\mathcal{E}_4(k)+\mathcal{O}(T^6)$. Similarly, we have $\bar Q=Q+T^2 Q_2+T^4 Q_4+\mathcal{O}(T^6)$. Performing the Sommerfeld-like expansion of Eq.~(\ref{eq:YY}) we obtain a hierarchy of equations for $\mathcal{E}(k)$, $\mathcal{E}_2(k)$, and $\mathcal{E}_4(k)$. They should be combined with Eq.~(\ref{eq:Energy-Q-condition}) as well as three equations obtained by successfully differentiating Eq.~(\ref{eq:LIE}) with respect to $Q$. There we account for the derivatives $\mathcal{E}^{(j)}(Q)$ and $\mathcal{E}_2^{(j)}(Q)$. The latter follow from
\begin{align}
\mathcal{E}_2(k) = -\frac{\pi^2 m}{3\hbar^2 n}\rho^{(0,1)}(k,Q),
\end{align}
and the former can be calculated using the formalism developed in Ref.~\cite{ristivojevic_method_2022}. The final results are
\begin{gather}\label{eq:eps0prime}
\mathcal{E}'(Q)={}\frac{\hbar^2 n}{2m}\frac{1}{\rho(Q,Q)},\\
\mathcal{E}^{(j+2)}(Q)={}\frac{2\pi\hbar^2 \rho^{(j,0)}(Q,Q)}{m}-\frac{\hbar^2 n \rho^{(j,1)}(Q,Q)}{2m\rho^2(Q,Q)},
\end{gather}
for $j=0,1,2,\cdots$. Eventually, using Eq.~(\ref{eq:pressureYY}) we can express the pressure as $\bar P=P+T^2 P_2+T^4P_4+\mathcal{O}(T^6)$, where $P=\bar \mu n-E_0/L$, $P_2=\pi/6\hbar v$. The expression for $P_4$ is significantly more complicated, see Appendix. It can be expressed in terms of $\rho(Q,Q)$ and its total derivatives, $\rho^{(1,0)}(Q,Q)$, and $\rho^{(2,0)}(Q,Q)$.
	
The pressure enables us to obtain the density $\bar n$ in the grand-canonical ensemble, $\bar n=({\partial\bar P}/{\partial\bar\mu})_T$. It depends on the temperature, $\bar n=n+T^2 n_2+\cdots$, with $n_{2}=(1/\hbar v)(\partial P_{2}/\partial Q)$. Here we have used $\partial\bar \mu/\partial Q=\hbar v$. In the final step we should transform the results to the canonical ensemble. There, $\bar n=n=N/L$ is fixed and $\bar\mu$ depends on the temperature, $\bar\mu=\mu+T^2\mu_2+\cdots$. Since $\bar\mu$, $n$, and $T$ are mutually related (the density is an integral of a function that depends on $\bar\mu$ and $T$ in the thermodynamic Bethe ansatz \cite{yang_thermodynamics_1969}) we use the exact relation 
\begin{align}
\left(\frac{\partial \bar\mu}{\partial T}\right)_n {\left(\frac{\partial n}{\partial \bar\mu}\right)_T}=-\left(\frac{\partial n}{\partial T}\right)_{\bar\mu}.
\end{align}
It enables us to find $\mu_2=-\pi\hbar v n_2/K$.
Substituting the low-temperature forms of $\bar P$ and $\bar\mu$, the free energy density takes the form
\begin{align}\label{eq:Fdensity}
\frac{F}{L}=\mu n-P-P_2T^2- \left(P_4+\frac{n_2\mu_2}{2}\right)T^4+\mathcal{O}(T^6).
\end{align}
In order to obtain Eq.~(\ref{eq:Fdensity}) we have used the relations
$P'(\mu)=n$, ${P}_2'(\mu)=n_2$, and $P''(\mu)={K}/{\pi\hbar v}$,
where the prime denotes the derivative with respect to $\bar\mu$ at $\bar\mu=\mu$. Note that the leading temperature corrections to $\bar{n}$ and $\bar{\mu}$ are sufficient to find the term proportional to $T^4$ in the free energy. Substituting $\mu_2$ and $n_2$, the final result for $f(\gamma,\tau)=F/N\epsilon$ is given by Eq.~(\ref{eq:f}) with
	\begin{align}\label{eq:C4general}
		C_4{}&=-\frac{64\pi^6 \rho^8}{45}\biggl[ 1+\frac{3n\rho^{(1,0)}}{8\pi\rho^3} -\frac{3n^2\dot\rho\rho^{(1,0)}}{32\pi^2\rho^6}-\frac{5n^2\rho^{(2,0)}}{128\pi^2\rho^5}\notag\\& +\frac{5n^2(\rho^{(1,0)})^2}{256\pi^2\rho^6} -\frac{n\dot\rho}{2\pi\rho^3} -\frac{n^2\dot\rho^2}{256\pi^2\rho^6} +\frac{17n^2\ddot\rho}{512\pi^2\rho^5}\biggr].
	\end{align}
Here we have suppressed the arguments $(Q,Q)$ after $\rho$ and its derivatives. Using Eqs.~(\ref{eq:K1})--(\ref{eq:K2K1}) we can express $\rho^{(1,0)}(Q,Q)$ and $\rho^{(2,0)}(Q,Q)$ in terms of $K_1$. Then Eq.~(\ref{eq:C4general}) becomes the result (\ref{eq:C4}).

\textit{Discussion.---}Once the free energy is evaluated, the calculation of other thermodynamic parameters in the canonical ensemble is simple. The chemical potential is given by $\bar\mu=(1/L){\partial F}/{\partial n}$ and the pressure is $\bar P=\bar\mu n-n\epsilon f(\gamma,\tau)$, the entropy is $S=-{\partial F}/{\partial T}$, and the internal energy is $U=F+TS$. Finally, the specific heat is given by $C={\partial U}/{\partial T}$. After evaluation it reads
\begin{align}
C(\tau)=N\left(\frac{K}{6}\tau-12C_4 \tau^3+\mathcal{O}(\tau^5)\right).
\end{align}
Therefore, $C_4$ controls the leading correction term. {Changing the sign as the interaction strength is increased, see Fig.~\ref{fig:C4}, the coefficient $C_4$ is directly responsible for the change of the specific heat behavior at a fixed temperature from concave at $\gamma<0.606$ to convex at $\gamma>0.606$. The curvature in the specific heat $C(\tau)$ would thus serve as a direct way to experimentally measure $C_4$.}
	
{The ground state of the Lieb--Liniger Bose gas has an appealing picture where the rapidities are densely packed in the Fermi-sea-like structure between the two Fermi rapidities \cite{lieb_exact_1963,korepin}. In many respects, the rapidities resemble the momenta of free fermions that fill the Fermi sea. The corresponding rapidity distribution, measured in experiments \cite{wilson_observation_2020,dubois_probing_2024}, is directly linked to the Luttinger liquid parameter, which follows from the value of the distribution at the edge, i.e., at the Fermi rapidity, see Eq.~(\ref{eq:Kll}). In this work, by Eq.~(\ref{eq:Kj}) we defined the family of parameters $K_j$  that parametrize the derivatives of the rapidity distribution at the Fermi rapidity. They thus have a clear intuitive meaning, characterizing the local behavior of the rapidity distribution near the Fermi rapidity. Interestingly, we showed that $K_j$ parameters are not independent, but hierarchically ordered, satisfying certain differential equations, which can be explicitly solved. We have performed this procedure here for $K_1$ and $K_2$.
}

{The family of parameters (\ref{eq:Kj}) is important well-beyond the Yang--Yang thermodynamics. One example is the low-momentum spectrum of elementary excitations \cite{petkovic_spectrum_2018}. For particles, the spectrum is given by $\varepsilon_p=v|p|+p^2/2m^*+\chi |p|^3/6\pi\hbar m n+\cdots$, where the sound (Fermi) velocity $v$ and the effective mass $m^*$ can be expressed in terms of $K$. The exact result for the cubic coefficient $\chi$, on the other hand, depends on $K$ and $K_1$. It is given by
\begin{align}
\chi=\frac{K_1}{K^2}+\frac{\gamma K'K_1}{2K^3} -\frac{\gamma K'}{2K} +\frac{\gamma^2 K'^2}{8K^2}- \frac{\gamma^2 K''}{4K}.
\end{align}
We note that the quartic coefficient of the spectrum can be expressed in terms of $K$, $K_1$, and $K_2$, where the latter can be reexpressed in terms of $K_1$ due to Eq.~(\ref{eq:K2K1}). Another example where the parameters $K_j$ will participate is the dynamical correlation function at low temperatures \cite{esler_temperature_1998}. Its evaluation is however beyond the scope of the present work. }

Until now we have been focused on the Lieb--Liniger model. However, many of the obtained results are valid beyond it. Equation (\ref{eq:LIE}) supplemented by a proper scattering phase shift also describes other Galilean-invariant integrable models. In the case of nonsingular phase shifts, the density of quasimomenta is an analytic function and Eq.~(\ref{eq:PDE}) will remain valid \cite{petkovic_spectrum_2018,ristivojevic_exact_2023}. 
Therefore Eqs.~(\ref{eq:Fho10}) and (\ref{eq:Fho20}) are unaffected. Moreover, Eq.~(\ref{eq:f}) is a general form of the dimensionless free energy per particle with $C_4$ given by Eq.~(\ref{eq:C4general}). A notable example of Galilean-invariant integrable models with noncontact interaction is the hyperbolic Calogero--Sutherland model. It is characterized by the interaction potential $V(x)=\hbar^2 \lambda(\lambda-1)\kappa^2/m\sinh^2(\kappa x)$ \cite{sutherland}. However, finding explicit results for the latter model is beyond the scope of this work as we are not aware of analytical expression for the Luttinger liquid parameter $K$ in terms of the microscopic parameters of the model.

In conclusion, we have developed the formalism that enables the treatment of thermodynamics of Galilean-invariant integrable models at low temperatures at arbitrary interactions. The obtained results are beyond the ones {calculated} using conformal field theory. The leading correction to the old result for the free energy is accounted by Eqs.~(\ref{eq:f}) and (\ref{eq:C4general}). It can be expressed in terms  of the Luttinger liquid parameter $K$ and newly introduced dimensionless parameters $K_1$ and $K_2$, see Eqs.~(\ref{eq:K1}) and (\ref{eq:K2}). For the special case of the Lieb--Liniger model, due to the connection (\ref{eq:K2K1}), only $K$ and $K_1$ are involved, see Eq.~(\ref{eq:C4}). Moreover, for the latter model we have explicitly calculated the corresponding series expansions at weak and strong interactions. Our approach can be extended to account for further terms in the free energy. {Since the rapidity distribution is experimentally measured \cite{wilson_observation_2020,dubois_probing_2024}, its local properties such as the derivatives at the edge described by $K_j$ parameters should also be amenable to experimental probes.}

Work at Laboratoire de Physique Th\'{e}orique was supported in part by the EUR grant NanoX ANR-17-EURE-0009 in the framework of the ``Programme des Investissements d’Avenir''.


%

\onecolumngrid	

\newpage
\begin{center}
{\large\textbf{End matter}}
\end{center}
\twocolumngrid	

\begin{table}[b]
\caption{The first two columns are the numerical values for pairs $\gamma$ and $C_4$ of the data points shown in Fig.~\ref{fig:C4}. The error in the numerical values is due to the rounding error on the last digit. The third column shows the relative difference between the exact value of $C_4$ and the analytical result (\ref{eq:C4small}) for small $\gamma$ denoted here by $C_4^{<}$. We list the values for $\gamma<11$ since Eq.~(\ref{eq:C4small}) is not valid for large $\gamma$. Seemingly small difference even for $\gamma$ larger than $10$ is an artifact of the accuracy of Eq.~(\ref{eq:C4small}). It does not contain the terms proportional to $\sqrt\gamma$ and to higher powers of $\sqrt\gamma$ that would cause rapid growth of $C_4^{<}$ at such large $\gamma$. The fourth column shows the relative difference between the exact value of $C_4$ and the analytical result (\ref{eq:C4large}) for large $\gamma$, denoted by $C_4^{>}$. We list the relative error for $\gamma>2$, since it grows significantly for smaller $\gamma$.\label{table1}}
\begin{ruledtabular}
\begin{tabular}{LLQQ}
\makecell[c]{\gamma} & \makecell[c]{C_4} & \makecell[c]{\left|\frac{C_4-C_4^{<}}{C_4}\right|} & \makecell[c]{\left|\frac{C_4-C_4^{>}}{C_4}\right|} \\
\hline
0.52963204 & 0.013625418 & 2.3\times 10^{-3} & - \\
0.54829049 & 0.0094180512 & 3.4\times 10^{-3} & - \\
0.56715152 & 0.0058146836 &  5.6\times 10^{-3} & - \\
0.57983606 & 0.0037043476 & 8.9\times 10^{-3} & - \\
0.59581397 & 0.0013551446 &  2.5\times 10^{-2} & - \\
0.62490962 & -0.0021780448 & 1.6\times 10^{-2} & - \\
0.66435552 & -0.0057655561 & 6.2\times 10^{-3} & - \\
0.70451958 & -0.0083715956 & 4.4\times 10^{-3} & - \\
0.74537414 & -0.010242963 & 3.8\times 10^{-3} & - \\
0.78689291 & -0.011561895 & 3.4\times 10^{-3} & - \\
0.82905087 & -0.012464020 & 3.3\times 10^{-3} & - \\
1.0000000 & -0.013601469 & 3.4\times 10^{-3} & - \\
1.2810553 & -0.012379481 & 4.5\times 10^{-3} & - \\
1.4748293 & -0.011147225 & 5.5\times 10^{-3} & - \\
1.7254056 & -0.0096739364 & 7.1\times 10^{-3} & - \\
2.0000019 & -0.0083322398 & 9.3\times 10^{-3} & 5.3 \\
2.5204513 & -0.0064742918 & 1.5\times 10^{-2} & 1.9 \\
3.0000213 & -0.0053139725 & 2.1\times 10^{-2} & 8.7\times 10^{-1} \\
4.0000598 & -0.0038307778 & 3.6\times 10^{-2} & 2.2\times 10^{-1} \\
4.9999456 & -0.0029953234 & 5.5\times 10^{-2} & 7.2\times 10^{-2} \\
5.9991022 & -0.0024740240 & 7.6\times 10^{-2} &  2.7\times 10^{-2} \\
6.9963257 & -0.0021238858 & 9.7\times 10^{-2} & 1.1\times 10^{-2} \\
7.9895883 & -0.0018756047 & 1.2\times 10^{-1} & 4.5\times 10^{-3} \\
9.0005032 & -0.0016882465 &1.4\times 10^{-1} &1.9\times 10^{-3} \\
10.000121 & -0.0015463084 &1.6\times 10^{-1} &8.2\times 10^{-4} \\
11.000033 & -0.0014343546 & - & 3.2\times 10^{-4} \\
12.000010 & -0.0013440823 & - & 9.5\times 10^{-5} \\
13.000003 & -0.0012699263 & - & 4.8\times 10^{-6} \\
14.000001 & -0.0012080401 & - & 4.6\times 10^{-5} \\
15.000000 & -0.0011556892 & - & 5.9\times 10^{-5} \\
16.000000 & -0.0011108820 & - & 6.0\times 10^{-5} \\
17.000000 & -0.0010721360 & - & 5.5\times 10^{-5} \\
18.000000 & -0.0010383272 & -& 4.9\times 10^{-5} \\
19.000000 & -0.0010085890 & - & 4.2\times 10^{-5} \\
20.000000 & -0.00098224323 &- & 3.6\times 10^{-5} \\
\end{tabular}
\end{ruledtabular}
\end{table}

\setcounter{equation}{0}
\renewcommand{\theequation}{A\arabic{equation}}

\textit{Appendix A: Differential equations for $\rho^{(3,0)}(Q,Q)$ and $\rho^{(4,0)}(Q,Q)$.---}In order to illustrate the general procedure of obtaining the derivatives of $\rho^{(j,0)}(Q,Q)$ described in the main text, here we report the differential equations for the cases $j=3$ and $j=4$. These quantities would be important, for example, to obtain higher order corrections in the free energy at low temperatures. The final results are 
\begin{align}
\frac{d}{dQ} &\left( \frac{\rho^{(3,0)}(Q, Q)}{\rho(Q,Q)}\right) = \frac{R^2(Q)}{2} + \frac{1}{4} \frac{d^2 R(Q)}{dQ^2}\notag\\
& + \frac{1}{2} \frac{d R(Q)}{dQ} \frac{\rho^{(1,0)}(Q, Q)}{\rho(Q,Q)} +R(Q) \frac{\rho^{(2,0)}(Q, Q)}{\rho(Q,Q)}
\end{align}
and
\begin{align}
\frac{d}{dQ}& \left( \frac{\rho^{(4,0)}(Q, Q)}{\rho(Q,Q)}\right) = \frac{1}{8} \frac{d^3 R(Q)}{dQ^3} + \frac{1}{2} \frac{d R^2(Q)}{dQ}\notag\\
&+\left(\frac{R^2(Q)}{2} + \frac{1}{4} \frac{d^2 R(Q)}{dQ^2} \right) \frac{\rho^{(1,0)}(Q, Q)}{\rho(Q,Q)}\notag\\
&+ \frac{1}{2} \frac{d R(Q)}{dQ} \frac{\rho^{(2,0)}(Q, Q)}{\rho(Q,Q)}+R(Q) \frac{\rho^{(3,0)}(Q, Q)}{\rho(Q,Q)}.
\end{align}
We have also derived the equations for $\rho^{(5,0)}(Q,Q)$ and $\rho^{(6,0)}(Q,Q)$. They involve an increasingly larger number of terms and are thus not reported here.

{
\textit{Appendix B: Details about the numerical results.---} The numerical results for $C_4$ are calculated by solving the integral equations for $\rho(k,Q)$ and its derivatives entering Eq.~(\ref{eq:C4general}). We used the method developed in Ref.~\cite{ristivojevic_conjectures_2019} that leads to very precise results. For example, if needed, the precision of fifty or more digits in the numerical results can be obtained quickly on a personal computer. In Table \ref{table1}, the data used to plot Fig.~\ref{fig:C4} are given together with the relative error between the numerical results and analytically evaluated expressions (\ref{eq:C4small}) and (\ref{eq:C4large}) for $C_4$ in the limiting cases of weak and strong interactions.
}

\begin{figure}[t]
\includegraphics[width=0.95\columnwidth]{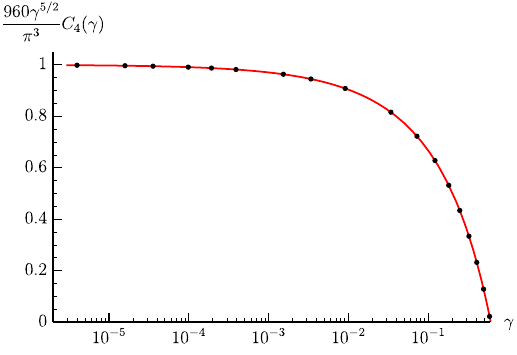}
\caption{Plot of the rescaled coefficient $C_4$ at weak interactions. The dots represent numerically exact values and the curve is obtained from Eq.~(\ref{eq:C4small}).}\label{fig:C4small}
\end{figure}

{
In the regime $\gamma\ll 1$, the coefficient $C_4$ diverges as $\gamma^{-5/2}$. In order to represent its behavior at very small $\gamma$ it is convenient to plot the rescaled quantity $960\gamma^{5/2}C_4/\pi^3$ that reaches $1$ when $\gamma\to 0$. This is shown in Fig.~\ref{fig:C4small} together with the analytical result (\ref{eq:C4small}). Over many orders of magnitude of $\gamma$, the agreement is perfect, see Table \ref{table2}. This is manifested by the relative error that scales with $\gamma$ as $\gamma^3$, which is in accordance with the leading neglected term in Eq.~(\ref{eq:C4small}). 
}
\begin{table}
\caption{The first two columns are the numerical values for pairs $\gamma$ and $C_4$ of the data points shown in Fig.~\ref{fig:C4small} obtained by the numerical solution of the problem. The error in the numerical values is due to the rounding error on the last digit. The third column shows the relative difference between the exact value of $C_4$ and the analytical result (\ref{eq:C4small}) for small $\gamma$, denoted by $C_4^{<}$. Note that for the third column are required the numerical values of $\gamma$ and $C_4$ that are more precise than the ones displayed in the first two columns.}\label{table2}
\begin{ruledtabular}
\begin{tabular}{LLQ}
\makecell[c]{\gamma }&\makecell[c]{\frac{960\gamma^{5/2}}{\pi^3}C_4} &\makecell[c]{ \left|\frac{C_4-C_4^{<}}{C_4}\right|  }\\
\hline
3.9875177\times 10^{-6} & 0.99814465 & 7.0\times10^{-20} \\
1.5907359\times 10^{-5} &  0.99629139 & 4.5\times 10^{-18} \\
3.5701738\times 10^{-5} &  0.99443977 & 5.1\times 10^{-17} \\
9.8704197\times 10^{-5} & 0.99074053 & 1.1\times 10^{-15} \\
0.00019260030 & 0.98704567 & 8.0\times 10^{-15} \\
0.00039057769 & 0.98150995 & 6.7\times 10^{-14} \\
0.0015324501 & 0.96309617 & 4.2\times 10^{-12} \\
0.0033880438 & 0.94471563 &  4.6\times 10^{-11} \\
0.0091155081 & 0.90797572 & 9.4\times 10^{-10} \\
0.034025401 & 0.81578647 & 5.6\times 10^{-8} \\
0.072091012 & 0.72258520 &  6.1\times 10^{-7} \\
0.12139633 & 0.62800326 & 3.4\times 10^{-6} \\
0.18044985 & 0.53181477 & 1.3\times 10^{-5} \\
0.24804839 & 0.43386053 & 4.4\times 10^{-5} \\
0.32319843 & 0.33401808 & 1.3\times 10^{-4} \\
0.40506558 & 0.23218708 &  3.7\times 10^{-4} \\
0.49293992 & 0.12828131 & 1.2\times 10^{-3} \\
0.58621104 & 0.022223843 & 1.2\times 10^{-2} \\
\end{tabular}
\end{ruledtabular}
\end{table}

\begin{figure}[t]
\includegraphics[width=0.95\columnwidth]{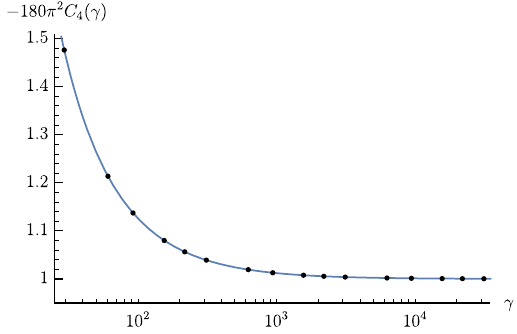}
\caption{Plot of the rescaled coefficient $C_4$ at strong interactions. The dots represent numerically exact values and the curve is obtained from Eq.~(\ref{eq:C4large}).}\label{fig:C4large}
\end{figure}

{ 
In the regime $\gamma\gg 1$, the coefficient $C_4$ approaches $-1/180\pi^2$. We thus plot $-180\pi^2 C_4$ as a function of $\gamma$ and study how it approaches $1$ as $\gamma$ is increased, see Fig.~\ref{fig:C4large}. We find excellent agreement with the analytical expression (\ref{eq:C4large}) over many order of magnitude, see Table~\ref{table3}. The relative error between the exact numerical result and analytical expression (\ref{eq:C4large}) decreases as $\gamma^{-6}$ with increasing $\gamma$, which is in agreement with the leading neglected term in Eq.~(\ref{eq:C4large}).
}

\begin{table}[b]
\caption{The first two columns are the numerical values for pairs $\gamma$ and $C_4$ of the data points shown in Fig.~\ref{fig:C4large} obtained by the numerical solution of the problem. The error in the numerical values is due to the rounding error on the last digit. The third column shows the relative difference between the exact value of $C_4$ and the analytical result (\ref{eq:C4large}) for large $\gamma$, denoted by $C_4^{>}$. Note that for the third column are required the numerical values of $\gamma$ and $C_4$ that are more precise of the ones displayed in the first two columns.}\label{table3}
\begin{ruledtabular}
\begin{tabular}{LLQ}
\makecell[c]{\gamma} & \makecell[c]{-180\pi^2C_4} &\makecell[c]{ \left|\frac{C_4-C_4^{>}}{C_4}\right| } \\
\hline
29.429050 & 1.4757438 & 7.4\times 10^{-6} \\
60.835173 & 1.2133648 &1.8\times 10^{-7} \\
92.249258 & 1.1371086 & 1.8\times 10^{-8} \\
155.08017 & 1.0798699 &9.0\times 10^{-10} \\
217.91176 & 1.0563303 &1.2\times 10^{-10} \\
312.15940 & 1.0390572 & 1.5\times 10^{-11} \\
626.31856 & 1.0193125 &2.4\times 10^{-13} \\
940.47781 & 1.0128273 &2.1\times 10^{-14} \\
1568.7963 & 1.0076736 &1.0\times 10^{-15} \\
2197.1149 & 1.0054741 &1.3\times 10^{-16} \\
3139.5927 & 1.0038282 &1.6\times 10^{-17} \\
6281.1853 & 1.0019120 & 2.5\times 10^{-19} \\
9422.7780 & 1.0012742 & 2.2\times 10^{-20} \\
15705.963 & 1.0007643 &1.0\times 10^{-21} \\
21989.149 & 1.0005458 & 1.3\times 10^{-22} \\
31413.927 & 1.0003821 &1.6\times 10^{-23} \\
\end{tabular}
\end{ruledtabular}
\end{table}

\setcounter{equation}{0}
\renewcommand{\theequation}{C\arabic{equation}}

\textit{Appendix C: Subleading corrections to the pressure and the free energy.---}The subleading temperature correction of the pressure in the grand-canonical ensemble is given by
\begin{align}\label{eq:P4final}
P_4=\frac{224\pi^6 m^3 \rho^8}{15\hbar^6 n^5}\biggl[ 1-\frac{13n\dot\rho}{21\pi\rho^3} +\frac{19n^2\dot\rho^2}{336\pi^2\rho^6} +\frac{17n^2\ddot\rho}{672\pi^2\rho^5}\notag\\
+\frac{2n\rho^{(1,0)}}{7\pi\rho^3} -\frac{n^2\dot\rho\rho^{(1,0)}}{14\pi^2\rho^6}  +\frac{5n^2(\rho^{(1,0)})^2}{336\pi^2\rho^6} -\frac{5n^2\rho^{(2,0)}}{168\pi^2\rho^5}\biggr].
\end{align}
As in the main text, we have suppressed the arguments $(Q,Q)$ of $\rho$ and its derivatives. Equation (\ref{eq:P4final}) arises from the Sommerfeld-like expansion of the pressure $\bar P$ and the pseudoenergy $\bar{\mathcal{E}}(k)$ combined with the relation of the latter with the ground-state distribution $\rho(k,Q)$ and its derivatives.

In the second temperature correction of the free energy (\ref{eq:Fdensity}), two terms appear. One is $P_4$ given by Eq.~(\ref{eq:P4final}) and the other is a combination $\mu_2 n_2$. The latter is related to the derivative of the sound velocity $v$ with respect to the Fermi rapidity $Q$. It can also be expressed in terms of the derivative of $v$ with respect to the chemical potential (at zero temperature). We have
\begin{equation}
	\mu_2 n_2 = -\frac{\pi^3}{36\hbar^3 K v}\left(\frac{\partial}{\partial Q} \frac{1}{v}\right)^2=-  \frac{m^3 K^2}{36 \hbar^4 n^3} \left(\frac{d v}{d \mu} \right)^2.
\end{equation}
Using $mvK = \pi \hbar n$ and $K=4\pi^2 \rho(Q,Q)^2$, we obtain
\begin{equation}
\frac{d v}{d \mu}  = \frac{4\pi \rho^2}{\hbar n} - \frac{2\dot{\rho}}{\hbar\rho}.
\end{equation}
Thus both contributions in term proportional to $T^4$ of the free energy can be expressed in terms $\rho(Q,Q)$ and its derivatives.

\end{document}